# *Lecture-Tutorials* in Introductory Astronomy

Colin S. Wallace and Edward E. Prather

**Abstract**

The *Lecture-Tutorials for Introductory Astronomy* have been designed to help introductory astronomy instructors actively engage their students in developing their conceptual understandings and reasoning abilities across a wide range of astrophysical topics. The development of the *Lecture-Tutorials* has been informed by nearly two-decades of research into common learning difficulties students experience when studying astronomy. The results from multiple studies provide evidence that *Lecture-Tutorials* can help students achieve learning gains well beyond what is typically achieved by lecture alone. Achieving such learning gains requires that an instructor understand how to effectively incorporate the *Lecture-Tutorials* into his or her course. This chapter provides details into the best practices for the effective integration and implementation of the *Lecture-Tutorials* – practices that we have developed through years of reflective practice from working with thousands of Astro 101 students and instructors. We also present a case study of how *Lecture-Tutorials* were used to promote the active engagement of learners in an Astro 101 mega-course enrolling over 700 students. This case study illustrates how the thoughtful implementation of *Lecture-Tutorials* can result in dramatic learning gains, even in the most daunting instructional environments.

**Introduction**

Imagine you have been assigned to teach a typical general education, college-level, introductory astronomy course (hereafter, Astro 101). A large number of students enroll in the course – perhaps as many as a few hundred, depending on your institution. All of these students attend the same lecture section, which meets 2-3 times per week for at least 50 minutes at a time. Students sit in seats that are bolted to the ground in a stadium-style layout that focuses their attention towards the front of the lecture hall. Unlike your colleagues teaching introductory physics, your Astro 101 course has no breakout recitation sections or labs. You have limited TA support, if any. The vast majority of your students are non-STEM majors who, put off by their previous science and math classes, plan to make Astro 101 the final science course they ever take. A large percentage of your students are freshmen who are just learning how to succeed at college. Many of your students may be first-generation college students and/or come from at-risk populations that have disproportionately high DFW rates in STEM courses. You know that there is a body of educational research that suggests that you should "actively engage" your students (Freeman *et al*., 2014), but the entire structure of the course seems designed to thwart any attempts to do anything other than just lecture at your students. What else could you possibly do?

Fortunately, there are a wide variety of pedagogical strategies that enable instructors to actively engage their students, even in the sub-optimal classroom environment described above. These strategies include Think-Pair-Share (also known as Peer Instruction;

Mazur, 1997), Ranking Tasks (Hudgins *et al*., 2007) and Interactive Lecture Demonstrations (Sokoloff and Thornton, 2001). In this chapter, we describe another instructional tool that Astro 101 instructors can incorporate into their curricula to actively engage students' learning: *The Lecture-Tutorials for Introductory Astronomy* (Prather *et al*., 2005; Prather *et al*., 2013). Each *Lecture-Tutorial* is a 2-7 page worksheet that addresses a single topic at a level appropriate for Astro 101 students. As described in more detail below, *Lecture-Tutorials* are designed to be completed in class by students working in collaboration with one or two of their peers. Each *Lecture-Tutorials* is a standalone activity that has been created to supplement lecture and, more importantly, be incorporated into class only after students have been taught about the topic of the *Lecture-Tutorial*; in this sense they should be thought of as a post-lecture activity. A single *Lecture-Tutorial* provides a sequence of questions that intentionally elicit and confront students' incorrect ideas (Clement, Brown, and Zeitsman, 1989), build upon their productive intuitions and beliefs (McDermott, 1991), and guide their learning to help them develop more expert-like understandings of astrophysical topics. *Lecture-Tutorials* often contain tables of data, figures, and other discipline specific representations that students must reason about. Many of these representations have been specially designed to foster the learning of a particular topic and may or may not correspond to typical textbook figures or to representations that are commonly used by experts in astrophysics; consequently, we refer to these specially created figures as *pedagogical discipline representations* (Wallace, Chambers, and Prather, 2016). The *Lecture-Tutorials* span the range of topics taught in Astro 101, from lunar phases and the seasons, to light and spectroscopy, to more advanced topics at the forefront of modern astrophysics, such as cosmology (Wallace, Prather, and Duncan, 2012), molecular excitations and synchrotron radiation (Wallace *et al*., 2016), and the detection of exoplanets via gravitational microlensing (Wallace, Chambers, and Prather, 2016). The development of each *Lecture-Tutorial* is informed by research into how people learn and common conceptual and reasoning difficulties experiences by Astro 101 students. Multiple studies show that students who use the *Lecture-Tutorials* significantly improve their understandings of the associated astrophysical topics beyond was is typically achieve by lecture alone (Prather and Brissenden, 2008; Prather *et al*., 2005; LoPresto and Murrell, 2009; Wallace, Prather, and Duncan, 2012).

Of course, the success of the *Lecture-Tutorials*, like any pedagogical tool, depends on how it is used. While research has shown that the use of active-learning strategies can help students achieve high learning gains, our research provides compelling evidence that how an instructor implements active learning is perhaps the most important factor in determining how much his or her students learn (Prather *et al*., 2009; Wallace, Chambers, and Prather, 2018). Since the early 2000s, we have engaged in classroom research, worked with thousands of faculty in professional development workshops, and reflected on our own practices. These experiences have helped us significantly evolved our implementation practices in order to maximize the effectiveness of the *Lecture-Tutorials*. In this chapter, we will unpack many implementation issues and solutions in the hopes that this discussion will help other Astro 101 instructors create more successful classroom environments. We will also present a case study of an Astro 101 "mega-course" enrolling over 700 students as a way of demonstrating how a series of thoughtful pedagogical decisions can result in an active learning classroom that uses *Lecture-Tutorials* to foster and support improved student learning.

Before we start our discussion, its critical to note that there are many more issues that one may encounter when implementing active learning strategies in the classroom than we can possibly unpack and work through in this article. While we do not wish to scare off instructors thinking of moving toward active learning, one needs to be ready to deal with how to orient students to the active learning classroom, how to motivate students to participate, how to establish classroom norms for collaborative groups, how to work with groups who are struggling or underperforming, how to monitor and provide real-time feedback to student on their success, and how to match course assessments to in-class experiences. For this manuscript we will address several key implementation issues that are specific to the effective use of the *Lecture-Tutorials*.

**Preparing to Implement *Lecture-Tutorials***

Our experience indicates that for the *Lecture-Tutorials* to be effective, students must believe that the *Lecture-Tutorials* will help them succeed on the course's homework, midterms, final exam, and other assessments. Consequently, instructors need to think carefully about how they will integrate the *Lecture-Tutorials* (and any other pedagogical tool or strategy for that matter) into the different elements of their course. For example, the *Lecture-Tutorial* "Telescopes and Earth's Atmosphere" focuses on developing students' understandings of what types of electromagnetic radiation penetrate Earth's atmosphere and why some telescopes are located in space and others on mountaintops. Imagine an Astro 101 instructor gave a lecture on ray optics, different types of telescopes (e.g., Newtonian vs. Cassegrain), and optical effects such as chromatic aberration, and then assigned this *Lecture-Tutorial* to his or her students. Even though both the lecture and the *Lecture-Tutorial* are superficially about "telescopes," the lecture did very little to prepare students for the reasoning tasks presented in the *Lecture-Tutorial*, leading students to view the *Lecture-Tutorial* as "a waste of time" and "irrelevant," especially if homework and exam questions focus exclusively on the non-*Lecture-Tutorial* content from the lecture. This poor alignment between the *Lecture-Tutorials* and other components of the class will significantly degrade the effectiveness of using *Lecture-Tutorials* in the classroom.

Before adopting any piece of curricula, textbook, or pedagogical strategy, instructors should determine the learning outcomes for the course, decide what evidence they need to determine whether those outcomes have been met, design the assessments needed to collect that evidence, and then assemble a curriculum that will prepare students for those assessments (Wiggins and McTighe, 1998). This process of "backwards design" can – and should – be iterative, especially as an instructor gains experience with the course. An instructor who is trying to figure out what is reasonable to expect out of Astro 101 students can leverage the fact that many years of astronomy education research have produced valid and reliable assessments of students' understanding of light and spectroscopy, star properties, Newtonian gravitation, and cosmology, to name just a few topics (Bailey *et al*., 2012; Bardar *et al*., 2007; Wallace, Prather, and Duncan, 2011;

Williamson, Willoughby, and Prather, 2013). There are also classroom-tested questions that can be used for Think-Pair-Share (TPS) and/or for assessments that are available from the Center for Astronomy Education (https://astronomy101.jpl.nasa.gov/materials/). Perusing these TPS questions will illustrate that Astro 101 students are capable of engaging in sophisticated critical thinking that goes well beyond simple regurgitation of facts, as these questions involve multistep reasoning, and the interpretation and integration of complex astronomical scenarios, numerical ideas, and cognitive tasks. Engaging in the process of backward design helps an instructor set realistic yet challenging learning outcomes for his or her students, and it helps the instructor make sure that all aspects of the course – its goals, its assessments, and its curricula – are aligned with one another.

We will now work through an example to illustrate how the principles of backward design can be applied to the teaching of Astro 101 using *Lecture-Tutorials*. First, imagine that you wish to get your students to understand how we use observations of Doppler-shifted starlight in the process of detecting extrasolar planets. Figure 1 shows a question that we often use to assess our own Astro 101 students' understanding of this topic. Consider all the discipline ideas and relationships students must coherently use in order to answer this question and the intellectual effort required to unpack its representations. They must have a fundamental understanding of what "radial velocity" is. They must recognize that it is the star's radial velocity that is provided on the graph, not the planet's (despite the fact that this is explicitly stated in the stem of the question, it is still a common source of difficulty for many students and faculty). They must remember that negative radial velocities correspond to a star whose light is blueshifted due to its motion toward Earth and positive radial velocities correspond to a star whose light is redshifted due to its motion away from Earth. Students need to be able to take the orbital information coded in the instant identified on the star's radial velocity graph and correctly translate that into the correct position in the diagram of the star's orbit. They must understand that both the star and the exoplanet orbit a common center of mass and that at any given time the star and planet are located on opposites sides of this center of mass – which requires that the star and planet are always moving in opposite directions and complete their orbits in the same amount of time. Students must be able to correctly interpret the picture of the orbits of the star and planet and determine how both are moving and where both are located at any instant in time relative to the identified location of Earth. This is a conceptually rich and challenging question whose surface features (e.g., the location of Earth, the time of the dot on the radial velocity curve, the direction of the star's motion, etc.) can be readily altered in order to create a large number of variants that are equivalent cases to the one shown but are *prima facie* novel to students. However, as we can attest from years of classroom instruction, it is not beyond the capabilities of Astro 101 students to correctly answer this question – but only if they have received ample time and intellectual engagement in order to develop their mental models on this topic beyond what is achieved from lecture alone.

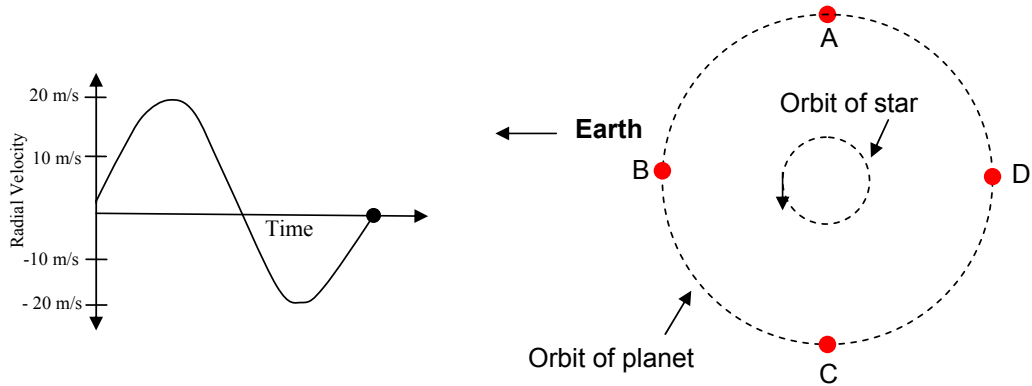

Figure 1. A sample question that probes students' understanding of Doppler shift and the detection of extrasolar planets from the radial velocities of their stars.

For the sake of our example, now imagine you are an Astro 101 instructor who wants to use the question in Figure 1 because it aligns with and assesses one or more of your course learning outcomes. You must consider how you will adequately prepare your students to be successful on a question like this. The *Lecture-Tutorial* "Motion of Extrasolar Planets" is specifically designed to help students develop the disciple knowledge, conceptual understandings, and reasoning abilities needed to answer the question in Figure 1. Additionally, students will probably benefit from a robust understanding of the Doppler shift, which is addressed in its own eponymous *Lecture-Tutorial*.

At this point, we must again emphasize that the *Lecture-Tutorials* are not designed to be stand-alone activities that students engage with independent of other elements of the course; rather, they are meant to be done in-class following a very targeted lecture on the topic. So, after deciding on your course learning outcomes, and the corresponding assessments, and after choosing the *Lecture-Tutorials* that will help students achieve those outcomes and perform well on the assessments, an instructor must construct classroom experiences that prepare students for the tasks contained in the selected *Lecture-Tutorial*. While we will refer to these classroom experiences as "lecture," since they typically take place in the lecture setting of an Astro 101 course, we do not mean to imply that an instructor should only be lecturing during this time.

There are two main goals for the lecture. First, the instructor must provide his or her students with the foundational knowledge they need as well as opportunities for those students to practice applying that knowledge. Second, the instructor must solicit from and provide to students feedback on how their understanding of the material is progressing. This sort of feedback is commonly referred to as formative assessment, since it is provided not for the purpose of assigning grades (although some instructors do award a small amount of participation credit) but rather the intent is to help students gauge their progress; research shows that the metacognition afforded by formative assessment serves as one of the most powerful ways an instructor can improve student learning and close pernicious achievement gaps between students from different demographic groups (Black and Wiliam, 1998).

In order to accomplish these two goals, an instructor should mix the traditional information-delivery aspect of lecture with instructional strategies that both actively engage students in developing their mental models on the topic and provide formative feedback. One effective active engagement instructional strategy we've implemented to supplement instruction is "50/50" and "Fill in the Blank" interactive lecturing. As an example, after a brief background lecture of the Doppler effect, a lecturer using the "50/50" and "Fill in the Blank" method could start off by telling the class "When the star is moving toward Earth, its light will be", and then pause. Students then shout out, simultaneously, "blueshifted." Alternatively, the instructor can provide a pair of choices for students, e.g. "When the star is moving toward Earth, will its light be blueshifted, or redshifted?", and then pause. This interactive lecture could continue with "Will the features in the spectrum for a star whose light is blueshifted appear at longer or shorter wavelengths?", and then pause. This "50/50" and "Fill in the Blank" interactive lecturing technique can be used multiple times during a lecture. The sequenced questions offers a quick and effective way for the instructor and students to gain real time feedback on whether the students are building a coherent model on the foundational knowledge that is essential for being able to reason about the topic being studied. This interactive lecturing technique needs to be established as a class norm at the start of the semester. We inform students that during lecture there will be times when they know the next word or right answer to a 50/50 statement, and that we simply want the entire class to say their answer out loud when we pause. We have found this to be an incredibly effective method for helping students organize their knowledge, for providing feedback, and for motivating students to engage with the topic.

In order to evaluate whether the majority of students have developed the discipline understanding necessary to engage with the Lecture-Tutorial we recommend instructors incorporate a series of Think-Pair-Share (TPS) questions into their lecture. It's important that these TPS questions address the different representations, conceptual ideas, and reasoning abilities critical to doing the *Lecture-Tutorial*. Instructors who are interested in how to effectively implement Think-Pair-Share should consult the guidelines at https://astronomy101.jpl.nasa.gov/download/workshopfiles/Think-Pair-ShareHow-ToGuide.pdf. For a bank of TPS question that align with the Lecture-Tutorials, instructors should visit https://astronomy101.jpl.nasa.gov/materials/. The feedback provided by effective use of TPS questions can help an instructor to adjust his or her instruction in real-time before starting the *Lecture-Tutorial*. More important, these TPS questions help students see how the information presented in lecture, covered in the *Lecture-Tutorial*, and featured on the course exams are all connected to one another. This awareness serves as a strong motivator for students to earnestly engage with the *Lecture-Tutorial*.

Once students have experienced an interactive lecture and responded to TPS questions on the key discipline ideas, we ask students to collaborate with their classmates on the *Lecture-Tutorial*. Every time you have students do a *Lecture-Tutorial*, we recommend that you provide them with the following instructions (shown in italic):

- *Work with a partner*. The *Lecture-Tutorials* are meant to be collaborative activities, so students should work with one or two of their classmates. Discourage groups of more than 3 students, since not every student may fully participate in larger groups. This is especially true for students located at the end of a row of several students who are working together.
- *Read the instructions and the questions carefully*. Many student questions are asked because students often fail to read the content of the LT question carefully enough to understand what is being asked. As we describe in the next section, whenever a student asks about a question, your first response should be for the student to read you the question in its entirety.
- *Discuss the concepts and your answers with each other. Take the time to understand the material now. It will help you on the exam.* Reinforce to students that the *Lecture-Tutorial* is not disconnected from the rest of the course; on the contrary, it is a critical part of their preparation for the exams. We often tell students that, by writing out their answers to the *Lecture-Tutorial*'s questions, they are authoring their own textbook – so they should be the best authors they can possibly be! Of course, this instruction will lose its force if your exams do not assess the concepts, skills, and reasoning abilities developed by the *Lecture-Tutorials*, so make sure that your exams do accurately and adequately reflect *Lecture-Tutorial* content.
- *Come to a consensus on your answer before your group moves on to the next question*. Since the *Lecture-Tutorials* are collaborative activities, students must actually work together to complete them. We have conducted research that strongly suggests that student learning during a *Lecture-Tutorial* is driven by the discourse they have with their classmates as they defend and debate their answers (Eckenrode, Prather, and Wallace, 2016). We cannot emphasize enough how important it is for students to participate in lively discussions of their reasoning while working through the *Lecture-Tutorial*. Note that a significant number of questions in a given *Lecture-Tutorial* will end with the phrase "explain your reasoning" to encourage these discussions. It is critical for faculty to set a high standard for student participation and level of discourse at the start of the course and continuously throughout the course. You can, and should, let your entire class and individual groups know when their level of engagement is not meeting your expectation. One easy to implement way to accomplish this is to tell the class that "it's too quiet" while they are working on their *Lecture-Tutorial*. When a particular low group-performing group asks for help, this is also a good time to remind them of your expectation of their level of discourse.
- *If you are stuck or not sure of your answer, check with a nearby group*. While we discourage groups of more than 3 students, different groups are certainly welcome to interact with one another. This furthers the opportunities for students to engage in discussion and debate with their peers, and it reduces the over-reliance of students on the instructor to answer all questions that might arise.

- *If you are really stuck or don't understand what the tutorial is asking, raise your hand and ask for help.* While students are working on the *Lecture-Tutorial*, the instructor and teaching assistance should be circulating around the room, listening to students' conversations, and responding to questions as they arise. For advice on how to address students' questions without re-lecturing to them, see the following section. If you are worried about having a student-to-instructor ratio that is too large for you to efficiently respond to every group with a question, read the case study below on teaching a mega-course and how we used undergraduate teaching assistants to help offset this ratio issue.

We recommend projecting these instructions (shown in italic in the above bulleted list) on a PowerPoint slide that remains visible during the entire time students are working on the *Lecture-Tutorial*. You should also give these instructions verbally when students are ready to start their very first *Lecture-Tutorial* of the term. While you will not need to verbally repeat these instructions every time students do a *Lecture-Tutorial*, we have found that we usually need to re-emphasize some or all of these points several times throughout the semester.

After the time for the *Lecture-Tutorial* ends, we recommend that instructors spend some time debriefing the activity with students. Provide a few minutes for students to ask questions and get help with any conceptual or reasoning difficulties they experienced. You will probably need to help your students understand how to ask thoughtful questions such as "Can you clarify how the slope on a Hubble plot tells us whether the expansion rate of the universe is increasing or decreasing?", rather than just seeking a solution statement through a questions like "What is the answer to question 18?" You may also want to quickly debrief the specific difficult or interesting answers to the *Lecture-Tutorial*'s questions with the entire class at once using the "50/50" and "Fill in the Blank" methods mentioned above. We have found that including a thoughtful and targeted debrief plays a critical role in improving our students' attitudes and beliefs about doing the *Lecture-Tutorials*. Providing students the opportunity to ask question, get clarification and feedback on whether their answers are correct, and whether they are prepared for exam questions, has a profound effect on students' willingness to participate in working collaboratively on the *Lecture-Tutorials*, and more importantly, on their self efficacy (Bailey *et al.*, 2017).

With all of the suggestions about implementation given above, understanding about how to manage time is perhaps the most challenging issue that face instructors who wish to implement *Lecture-Tutorials,* and so we will take a moment to discuss this issue. We strongly recommend starting each *Lecture-Tutorial* by telling students how long they have to work on the activity. We recommend estimating 5-8 minutes per page, depending on the number of questions and complexity of the reasoning involved. You can always increase or decrease this amount of time based on how fast most students are progressing through the *Lecture-Tutorial*; however, it is important to give this time constraint so that students realize they must get to work and work efficiently, since they do not have an indefinite amount of time. Along these lines, you should keep track of when most students have completed a page, and then, when appropriate, say out loud to the class "if you are still on the first, second, etc., page, you are starting to fall behind." This feedback can be very helpful for groups, especially slower ones (including those who are not meaningfully engaging in the activity) to self-assess their efforts. When the vast majority of students are on the last page, we recommend asking students to "raise your hand if you are on the last page or are done", and then tell the class "you only have a few more minutes to go." By having a clear majority of students raise their hands, you are communicating to <u>all</u> students that, most students were able to complete the LT in the afforded time, and that it is reasonable for you to wrap-up the *Lecture-Tutorial* period in the next couple minutes. Of course, there will always be some students who do not finish. Do not make them feel bad – especially if they were giving an earnest effort – but ask them to finish the *Lecture-Tutorial* outside of class – and invite them to work with you in your weekly "free help session" (typically referred to as office hours). In fact, its good practice to encourage all students to revisit the *Lecture-Tutorial* soon after class and take that time to better understand the questions, representations, tasks and answers for the *Lecture-Tutorial*. Even students who finished the activity may have only jotted down the "bare bones" of an answer and explanation to each question, so in the spirit of being good authors of their textbook, they should flesh out their explanations while the material is still fresh in their minds.

While you may only want to do only a few of *Lecture-Tutorials* the first time you use them in a course, we have found through experience that many students will only take the *Lecture-Tutorials* seriously when they are a regular component of the course and when they are introduced early in the course (within the first 2 or 3 classes). When used only infrequently, the *Lecture-Tutorials* can seem like a strange addition to the course pedagogy; students are not completely sure of what the purpose of the activity is, what their behavior is supposed to be, how the activity relates to their overall grade and course success, and as a result they may not engage as fully with the *Lecture-Tutorials* activities as do students in courses where *Lecture-Tutorials* are a significant and regular part of the curriculum.

Since the *Lecture-Tutorials* require approximately 10-20 minutes of class time, you may be worried about whether or not you actually have enough time in your class schedule to implement *Lecture-Tutorials*. This is where a thoughtful consideration of your goals for the class will help you decide how to efficiently budget your class time. For example, you may realize that you do not need to spend 50 minutes giving a traditional lecture on Kepler's $2^{nd}$ law. Instead, you may be able to accomplish your astronomical content goals by giving a tightly focused and streamlined 15-minute interactive lecture (including 50/50 and TPS questions) on Kepler's $2^{nd}$ law that helps students develop critical conceptual ideas and reasoning abilities, followed by 20 minutes on the associated *Lecture-Tutorial*. Additionally, when you take a more holistic look at your class schedule for the entire term, you may realize there are topics that you can de-emphasize or remove completely in order to make time for the *Lecture-Tutorials* on the topics you want to go deeper with and really engage your students about. Perhaps you have always devoted a lecture to planetary rings, only to retrospectively realize that you never cared enough about the topic to actually ask students meaningful questions about this topic on your exams. If it happens that a day spent on planetary rings does nothing to advance your goals for the course, then give yourself the freedom to drop the topic entirely. Alternatively, you may realize that rings are in fact important for your course goals, which may give you the impetus to develop the instructional experiences necessary advance students' understandings and the assessment questions necessary to measure whether students have actually met your learning outcomes for this topic. Either way, a holistic look at how you use your time with students may allow you to develop a more cohesive and focused learning experience for your students.

*Lecture-Tutorials* can be a powerful pedagogical tool and they can help your students significantly improve their conceptual understandings and reasoning abilities over a wide range of discipline content. However, they should not be simply dropped into a course with little thought about how they will be implemented and how they align with the rest of the course. Research shows that an instructor's ability to effectively implement active learning is perhaps the most important factor in determining the learning gains of his or her students (Prather *et al.*, 2009; Wallace, Chambers, and Prather, 2018). *Lecture-Tutorials* are no exception. An instructor adopting the *Lecture-Tutorials* needs to carefully consider all the issues related to their implementation described in this section. That being said, instructors should not be afraid to try the *Lecture-Tutorials* in their classes. Most students appreciate when their instructor has set clear goals and is utilizing activities designed to explicitly help them learn. As with any active engagement strategy, one's implementation of the *Lecture-Tutorials* gets better with practice. If you would like to get guidance and feedback on your implementation, consider attending one of the Center for Astronomy Education's professional development workshops (https://astronomy101.jpl.nasa.gov/workshops/) and/or interact with other Astro 101 instructors via the Astrolrner@CAE Yahoo Group (https://groups.yahoo.com/neo/groups/astrolrner/info).

**Best practices when facilitating collaborative groups working though *Lecture-Tutorials***

Inevitably, students will have questions as they work on a *Lecture-Tutorial*. In fact this is a good thing, and something that the *Lecture-Tutorials* were designed to promote, as students who are asking questions are students who are in a teachable moment. Be sure to constantly circulate around the room during the *Lecture-Tutorial* time. We find that students are more likely to ask questions when they see you nearby. When students ask a question, you must be ready to respond efficiently; you cannot spend too much time with an individual group, since you will probably have other groups of students with their own sets of questions. Your interactions with student groups must also keep the students actively engaged in the process of constructing their own understandings of the material. For these reasons, we strongly recommend that you not re-lecture information to students. Remember, the *Lecture-Tutorials* are post-lecture activities, so students have already heard your lecture on the relevant topic. Repeating the words you said earlier is probably not the most effective or efficient approach. Occasionally, you may find a group of students who somehow missed a key piece of knowledge from the lecture. If there is no way students could figure out that piece of information via asking them a set of 50/50 clarification questions, then you may need to just tell them a piece of information – but be concise, and resist going into full lecture mode. Students need for help, and the kinds of questions they will ask while doing their *Lecture-Tutorial* can come in many forms. We will now discuss several different methods we use to handle common interactions we have with student groups.

Students will often ask if their answer is correct. There are different ways you may want to respond to such a question. If the group is correct, and the question their inquiring about is relatively simple, then it is perfectly fine to just tell the students they are correct, and move on, as time is precious. However, for more difficult sophisticated questions that will typically also ask students to explain their reasoning, it's important to ask a group member (and we suggest asking the student who appears least engaged from the group) to explain their reasoning for their answer – and only affirm their correct answer if they can provide a correct explanation. If they cannot provide a correct explanation and/or if their answer is incorrect, then be prepared to engage them in a series of questions as described in the following examples.

Most student requests for help will be along the lines of "We don't understand what is going on in this situation," or "What is this questions asking" or "We don't even know where to begin with this question." Whenever students have a question related to the content and/or how to proceed down a particular reasoning pathway, your goal is to use a Socratic-style questioning technique to help guide your students' thinking along the right path. The number one instructional move we recommend starting with is to ask your students to read the *Lecture-Tutorial* question out loud and verbatim. This serves two purposes. First, it helps you make sure you know exactly which question they need help with. Second, and more importantly, some students do not read the questions in the *Lecture-Tutorial* very carefully before they call you over, so by having them read it aloud to you we often find that someone in the group will figure out the answer to their question as its being read aloud.

If simply reading the question aloud is insufficient, you have several different facilitation strategies that you can use. One thing to keep in mind is that you are trying to efficiently diagnose where your students are struggling, and to get them back on the correct path, but again time is critical, so it is best to think of this as emergency triage and not surgery. You can explicitly ask students to describe what they find confusing about the given question or situation (and you should do this if they have not already made clear what the specific issue is beyond "We need help on number 5."). If your students are able to articulate the issue they are having it can often be useful to have them reflect on their answer to a previous question in the *Lecture-Tutorial* that establishes critical information needed for the question they are asking about. In doing so, you may find that these students have made a particular reasoning error with the previous question, and now it is clear how best to help these students. The most common and effective facilitation strategy we see adopted by instructors is to ask targeted 50/50 questions to both diagnose students' conceptual or reasoning issue and to direct their thinking along a productive path (e.g. "If two stars are the same size, then will the hotter star give off more energy or less energy?"). As you work though your *Lecture-Tutorials* before class (which you definitely need to do), it can be helpful to anticipate which conceptual or reasoning issues your students may have with a particular question, and consider which variables you may need to frame in your 50/50 questions. You may need to ask students which variables or physical characteristics are important, and how they are related, for a given situation (e.g., "What two pieces of information do I need in order to determine the strength of the gravitational force between any two objects?" or "Which feature of the blackbody curve is directly related to the object's temperature?"). Depending on the specific circumstances, you may find that you need to mix-and-match these different facilitation strategies in order to help a group of students get unstuck. Throughout this process, however, never lose sight of the fact that your primary roll is to be a guide who employs Socratic-style questioning rather than an authority who provides answers. By guiding students with questions, you are keeping them actively engaged in the material and holding them responsible for constructing their own knowledge, which decades of research shows is necessary for deep and long-lasting understanding to develop (Redish, 1994).

Sometimes you will find a group whose students are not exhibiting the types of collaborative behavior you expect during *Lecture-Tutorial* time. One student may be dominating and/or doing all of the work. One student may be completely disengaged and trying to hide his cell phone use. The students may not all be on the same question, which suggests that they are not coming to a consensus on their answers. They may not be writing down their answers and their reasoning. Confronting and addressing these kinds of un-productive behaviors can be uncomfortable for you and for them. However, it can be done in a positive way that expresses your pedagogical goals and expectations and/or demonstrates your care for their learning. Remind students that all members of their group need to be involved in collaborating and coming to a consensus on their answers, as it is the level of discussion and consensus forming they foster in their group that will determine the level of improvement in their understanding they will experience from doing the *Lecture-Tutorial* activity (Prather *et al*., 2005; LoPresto and Murrell, 2009; Eckenrode, Prather, and Wallace, 2016). To try and get all group members to participate, you may want to direct your help and interactions to the group members who appears disengaged, who are not recording their answer, or who are working on a different question than the one you were called over to help out with. Imagine these different facilitation statements/questions that might be employed for the different issues we have raised: "I'll comeback once each member of your group is working on this question," or "John, what did you write down for the question Maria is asking about?" or "Marco what do you think about the question that Rosa just asked?" By not automatically responding to the group member that is working ahead or the dominant group member, you can pull the other students into the interaction, and communicate that your expectation is that all group members are involved in collaborating and coming to consensus. We also take the time to reinforce the idea that they are the authors of their own textbooks, so they need to write down abbreviated but cogent answers/explanations in class, and then re-visit and evolve these explanations as outside of class work. By doing this, they will set themselves up well when it comes time to study their *Lecture-Tutorials* while preparing for the exam. These messages are important for students to hear and it is perfectly normal to have to repeat them to students throughout the term. No matter which form of inappropriate group dynamic you observe, it's important to curb unproductive student behaviors as soon as possible. If some groups are allowed to exhibit these behaviors unchecked, then that sends an implicit message to the rest of the class that these behaviors are acceptable, which can inadvertently undercut much of what you are trying to accomplish with your implementation of the *Lecture-Tutorials*.

**Case Study: The Astro 101 Mega-Course**

As described above, the *Lecture-Tutorials* are one of several pedagogical tools and strategies that one can use to actively engage students in an Astro 101 class. The effectiveness of the *Lecture-Tutorials* has been supported by multiple studies (Prather and Brissenden, 2008; Prather *et al*., 2005; LoPresto and Murrell, 2009; Wallace, Prather, and Duncan, 2012). In this section, we add another study to the literature on the effectiveness of *Lecture-Tutorials* by showing how the instructional model described above can be applied to an Astro 101 course with a mega-enrollment of several hundred students.

One of us (Prather) began teaching mega-courses at the University of Arizona (UA) in the spring of 2010. While many previous reports on "large-lecture" or "mega-courses" discuss the challenges of teaching 100-300 students (e.g., Kapp *et al*., 2011; O'Moore & Baldock, 2007; Thanopoulos, 2004), the UA mega-courses see enrollments of 700-1400 students. To put this in perspective, an instructor teaching a single mega-course may be responsible for educating 3-5% of the *entire* undergraduate population of the UA in a single semester.

The challenges associated with teaching a mega-course are immense. The first issue is finding a classroom that can accommodate 1000 students. These UA mega-courses were taught in Centennial Hall, the university's 2000-seat performing arts center, which is designed for theater, orchestral, and ballet performances (Figure 2). Not only are the seats in the performing arts center bolted to the floor and unable to rotate, they even lack desktops! The rows are packed so close together that an instructor cannot walk across a row to reach students who need help. The lighting is much dimmer than a normal classroom's. The instructor must lecture from a raised stage, where he or she is dwarfed by a gigantic screen upon which lecture slides are displayed. As with many Astro 101 courses, there are no breakout recitation or laboratory sections, and there are no prerequisites for enrolling, which means our student population is very representative of the entire UA undergraduate population – approximately 1 in 5 of which will drop out of college after just one year. This confluence of factors appears to be a "worse case" scenario for many instructors, yet we took up the challenges of trying to re-create in this environment the same high level of student collaboration, interactivity, and learning gains we had previously achieved in smaller class.

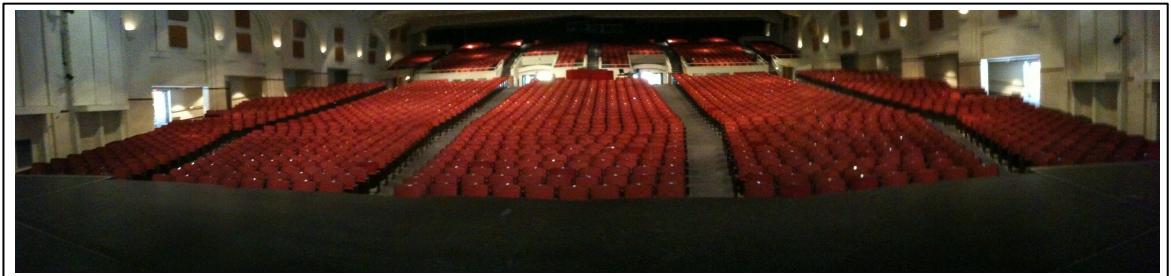

Figure 2. The view from the stage in Centennial Hall, the University of Arizona's performing arts center that also doubles as the classroom for the Astro 101 mega-course.

While the University of Arizona was motivated to create these mega-courses in response to budget and enrollment issues stemming from the Great Recession, we were interested to see if the results and lessons learned from the astronomy education research we had conducted with our colleagues over much of the preceding decade would translate well into an Astro 101 mega-course. Since the turn of the 21$^{st}$ Century, our research has uncovered many conceptual and reasoning difficulties that students experience with a wide variety of core topics in Astro 101.  The results of this work have been used to inform the development of pedagogical tools and strategies, including the *Lecture-Tutorials*, that explicitly address these difficulties.  At the same time, we've worked in collaboration with many astronomy education researchers to create several validated and reliable assessments that instructors can use to measure their students' understandings on a wide variety of Astro 101 topics (Bailey *et al*., 2012; Bardar *et al*., 2007; Wallace, Prather, and Duncan, 2011; Williamson, Willoughby, and Prather, 2013)  All of this work has been disseminated to faculty across that nation via the Center for Astronomy Education's Teaching Excellence Workshops (https://astronomy101.jpl.nasa.gov/workshops/).  These workshops have been designed to help faculty improve their abilities to effectively implement active learning (Prather and Brissenden, 2008).  Through iterative formative and summative research, reflective teaching practice in our classrooms, and by listening to the thousands faculty that have attended our workshops, we have significantly evolved our classroom practices, instructional models, and pedagogical training experiences – and we would need the best versions of all of these elements in order to engineer an effective active learning environment for the Astro 101 mega-course.

The instructional model described in the previous sections that combines interactive lectures, Think-Pair-Share, and *Lecture-Tutorials* was developed for and honed in classes that typically had enrollments of 150 - 200 students. While this enrollment is significantly smaller than the number of students in a single mega-course, 150 - 200 students is still a large number of students.  Because our prior research demonstrated that the *Lecture-Tutorials* and other aspects of this instructional model are successful in large classes, we had reason to believe that they would also work in a mega-course.

Several logistical issues had to be addressed in order to make the mega-course run smoothly. For example, as described earlier, the rows in this theater are packed so close together that we would not be able to get to students who needed help. To solve this access issue, we blocked off every fifth row of the class to prevent students from sitting in those rows. These empty rows allowed us to move quickly and easily throughout the room in order to have access to students who raise their hands for assistance during collaborative group work, or to quickly bring a student a microphone when he or she had a question during lecture. Many classroom practices that are simple and can be taken for granted become significant challenges in the mega-course, such as handing out and picking up classroom materials (such as participation forms, activity sheets, homework, and surveys). To deal with this, we formulated a intricate flowchart detailing exactly where in the room each teaching assistant should go and which rows they were to handout and pick up papers from so that all student could receive or submit their class materials in only a few minutes. Another difficulty with the mega-course was dealing class testing. For midterm examinations, we reserved Centennial Hall at evening times outside of normal class time. This meant that we were not using normal class time for testing, which allowed us to efficiently maintain exam security and check students' IDs, without having to worry about conflicting with other classes or scheduled performances. One aspect of this course that presented a unique challenge was how to accommodate office hours. Since between 10-30% of students in our courses typically attend office hours, we had to schedule a 150-seat classroom multiple times per week and staff it with two or three teaching assistants per meeting time. We also established and strictly enforced a zero tolerance cell phone and laptop policy in order to prevent hundreds of students from texting and using Facebook, YouTube, Twitter, or other websites that distract the learning of their fellow students. Studies have shown that students who spend class time using these electronic resources lower their grades by as much as a full letter grade (Duncan, Hoekstra, and Wilcox, 2012). Even if only 10-15% of our students use cell phones and laptops during class, that is still 75-110 illuminated screens in a dark room, which creates a serious distraction for the rest of the class. Students who legitimately wanted to take notes on their electronic devices were told they could do so in a specified area of the classroom. To our surprise, only approximately 10 students actually used their laptops and tablets for note taking during a given semester.

By far, the biggest issue we had to address for the mega-course was making sure we had enough instructional staff to deal with the significant increase in enrollment of this course. While all of the course's curriculum and assessments were already developed, and while a single individual could lead the lecture portion of the course, there was no way that one instructor and a single graduate teaching assistant (our normal ratio of instructional staff for a course of 150-200 students) could manage 700 or more students who were all working on a *Lecture-Tutorial* simultaneously. This is an issue that faces many Astro 101 instructors, who may have a large class and little or no TA support. While we were fortunate enough to have the Department of Astronomy provide two graduate teaching assistants for these courses, this was nowhere near an adequate number of instructors. Our solution: Hire approximately 8 former high-performing Astro 101 students who took the class in a previous semester to return to the classroom as peer teaching assistants (PTAs). As undergraduates, these former Astro 101 students are significantly cheaper than a graduate student, since they can be hired for an hourly wage and no benefits. Note that while our PTA program bears some similarity with the Learning Assistant (LA) program popular at many university physics programs (Otero *et al.*, 2006; Otero, Pollock, and Finkelstein, 2010), it differs from the LA program in that we recruited almost exclusively non-majors and we were not focusing our efforts on creating future K-12 STEM teachers. In these respects, our program was more similar to the Supplemental Instruction (SI) Program (Arendale 1997). Students hired as PTAs had to pass through an in-person interview in which they were placed in mock teaching scenarios, which required them to demonstrate their content understandings, communication abilities, and pedagogical abilities to effectively and efficiently help Astro 101 students. PTAs received specific training throughout the term focused on developing their pedagogical content knowledge (Gess-Newsome and Lederman, 1999). This was primarily done in weekly training meetings with the instructor, during which we reviewed the conceptual and reasoning difficulties associated with the coming week's *Lecture-Tutorials* and engaged in a version of situated apprenticeship (Prather and Brissenden, 2008) in which we modeled authentic student difficulties and mentored the PTAs in how to use Socratic-style questioning techniques to help student groups overcome those difficulties. Having a well trained cadre of PTAs proved to be absolutely essential for the successful implementation of the *Lecture-Tutorials* in the mega-course.

Our efforts paid off, as evidenced by the performance of mega-course students on the Light and Spectroscopy Concept Inventory (LSCI; Bardar *et al.*, 2007) as well as the Star Properties Concept Inventory (SPCI; Bailey *et al.*, 2012). In addition to assessing content that is fundamental to the course, the LSCI and SPCI also cover concepts and reasoning tasks that are addressed by multiple *Lecture-Tutorials*. In this section, we report on data from the mega-courses taught in the spring 2010, spring 2011, and spring 2012 (the mega-course was not offered in the fall semester). The SPCI was only given to students in the spring 2011. Other assessments, surveys and concept inventories were administered in other semesters for research projects that are beyond the scope of this chapter. Both the LSCI and SPCI were administered at the beginning and at the end of the semester so we could measure any learning gains achieved by students as a result of completing the course.

We only included data for students for which we had both their pre- and post-instruction responses. Furthermore, while we used Scantrons to collect our data in the spring 2010, in subsequent semesters we used an on-line password-protected system. Students received a nominal amount of participation credit for completing both their pre and post responses to the concept inventories, although their grades were not affected by their responses to individual questions. For the spring 2011 and 2012 semesters, we removed the responses of all students who spent fewer than 10 minutes or more than 60 minutes on the LSCI post-instruction. We also removed the responses of all students who spent fewer than 4 minutes or more than 60 on the SPCI. We looked at post-instruction times because by that point in the semester most students have developed the content knowledge to be able to really reason through the questions, as opposed to just guessing. Students who spend less than 10 minutes on the LSCI have less than 30 seconds, on average, to read, reason about, and answer each item, while students who spend an hour or more on one of these assessments are clearly taking much longer to respond to the twenty-six items than is intended. We found empirically that students are able to finish the SPCI in less time than the LSCI (with a median time of 8.5 minutes for the SPCI, compared to 15.2 for the LSCI), which is why we made the lower cut-off only 4 minutes instead of 10 minutes. By imposing these cut-offs, we removed 206 students with matched pre- and post-instruction responses from the spring 2011 LSCI data set, 74 students from the spring 2011 SPCI data set, and 117 students from the spring 2012 LSCI data set.

On the SPCI, the spring 2011 mega-course achieved an average normalized gain of $<g>$ = 0.39 with a standard deviation of $\sigma_{<g>}$ = 0.19 from $N$ = 307 matched pre- and post-instruction responses. This value of $<g>$ falls within Hake's 'medium gain' region (0.3 $\leq <g> < 0.7$; REF). To understand how the spring 2011 mega-course compares to other classes assessed with the SPCI, we compare our results to those presented in Bailey *et al.*'s (2012) validation study of the SPCI. The 334 Astro 101 students in Bailey *et al.*'s (2012) study had a pre-instruction average of 7.09 (with a standard deviation of 2.73) and a post-instruction average of 11.84 (with a standard deviation of 3.87). From these average scores, we calculate an average normalized gain of $<g>$ = 0.25. In contrast, the students in the spring 2011 mega-course had a pre-instruction average of 6.48 (with a standard deviation of 2.83) and a post-instruction average of 14.19 (with a standard deviation of 3.72). The fact that the students in the mega-course began with a lower pre-instruction average on the SPCI and ended with a higher post-instruction average than the students in Bailey *et al.* (2011) explains why the mega-course had the higher average normalized gain. These results indicate that students taught in an Astro 101 mega-course can significantly increased their understanding of topics related to star formation and stellar properties.

We now examine the results of all three mega-courses on the LSCI. Table 1 shows the average normalized gains, their standard deviations, and the number of students $N$ with matched pre- and post-instruction scores for each of the three semesters. Once again, all of these average normalized gains fall within Hake's 'medium gain' region.

Table 1. The average normalized gains $<g>$, their standard deviations $\sigma_{<g>}$, and the number of students $N$ with matched pre- and post-instruction LSCI scores for the spring 2010, spring 2011, and spring 2012 versions of the Astro 101 mega-course.

| Semester | $<g>$ | $\sigma_{<g>}$ | $N$ |
| --- | --- | --- | --- |
| Spring 2010 | 0.42 | 0.24 | 357 |
| Spring 2011 | 0.48 | 0.24 | 290 |
| Spring 2012 | 0.43 | 0.26 | 288 |

How do the average normalized gains of the mega-courses on the LSCI compare with the average normalized gains of Astro 101 courses across the country on the LSCI? In a previous study, we calculated the average normalized gains on the LSCI for sixty-nine classes, representing nearly 4000 students at thirty colleges and universities across the US, plus one in Ireland (Prather *et al.*, 2009). Out of those sixty-nine classes, only nineteen (27%) had average normalized gains in the 'medium gain' region – and only ten

(14%) of those classes had average normalized gains above 0.40. Because they all had average normalized gains above 0.40, the three mega-courses therefore achieved some of the largest learning gains of Astro 101 classes in the US, as measured by the LSCI.

We also re-analysed the LSCI data from Prather *et al*. (2009) plus the LSCI data from the mega-courses using item response theory (IRT). As described in more detail in Wallace, Chambers, and Prather (2018), IRT provides a way to estimate students' underlying abilities to reason about light and spectroscopy concepts, independent of the specific items to which they responded. Figures 3-5 show the distribution of student abilities, pre- and post-instruction, for the spring 2010, spring 2011, and spring 2012 mega-courses, respectively.

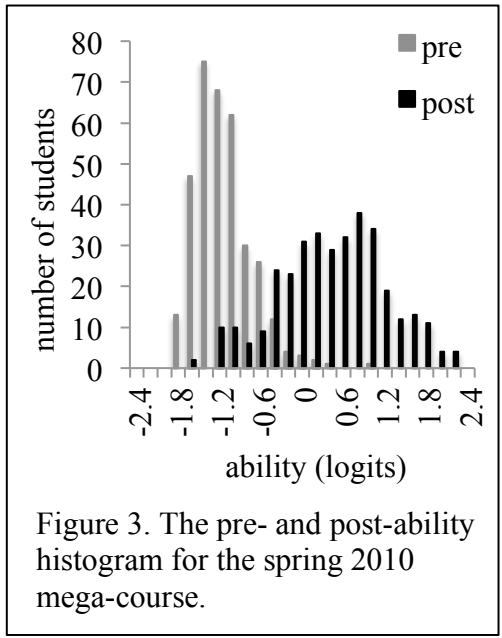

Figure 3. The pre- and post-ability histogram for the spring 2010 mega-course.

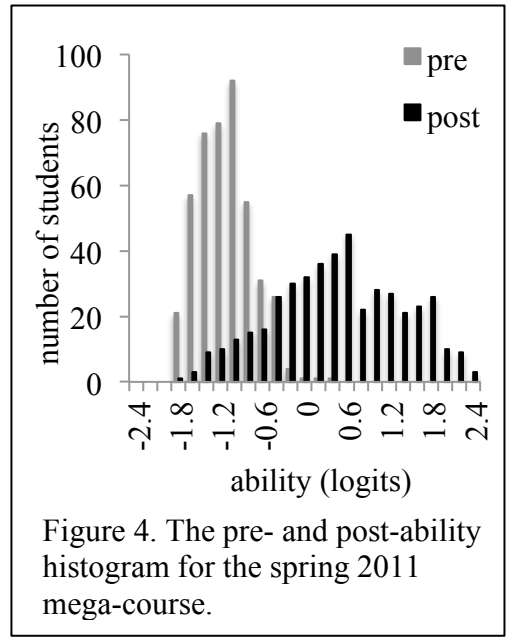

Figure 4. The pre- and post-ability histogram for the spring 2011 mega-course.

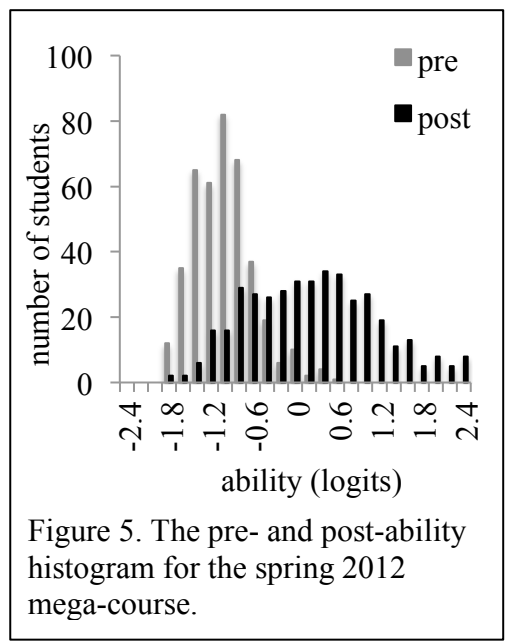

Figure 5. The pre- and post-ability histogram for the spring 2012 mega-course.

Note that in all three semesters there is a clear shift upwards in ability pre- to post-instruction. Post-instruction, the majority of students posses abilities that are far beyond what any student possessed pre-instruction. Additionally, the pre-instruction abilities represented in these three mega-classes are among the lowest in the entire national data set, whereas the post-instruction abilities are among some of the highest achieved by any student in the sample. This data provides powerful evidence that the proper implementation of *Lecture-Tutorials*, coupled with thoughtful solutions to the challenges facing active engagement in large-enrollment classes, can work together to significantly improve students' conceptual understandings and reasoning abilities, even in mega-courses containing several hundred students.

**Summary**

The *Lecture-Tutorials for Introductory Astronomy* are grounded in over 15 years worth of research into the conceptual and reasoning difficulties experienced by students taking an Astro 101 course. They can serve as a valuable tool to actively engaging students, even in classrooms that have hundreds of student groups operating simultaneously. But as is the case with all forms of active engagement, an instructor's implementation is critical. In this chapter, we have unpacked many of the instructional techniques we have developed over many years to facilitate the effective use of the *Lecture-Tutorials*. Instructors who are interested in using the *Lecture-Tutorials* should carefully reflect on how all aspects of the course – its goals, student learning outcomes, assessments, and curricula – fit together and mutually reinforce one another. As we have discussed throughout this manuscript, effective implementation of the *Lecture-Tutorials* is about more than just finding time for them in your class schedule. One must establish and communicate classroom and collaborative group norms, and adopt a backwards design approach to aligning assessments with all of the other aspects of the course, in order to establish a productive feedback loop between you and your students so that you can continuously work to foster an effective and vibrant active learning environment. While implementing the *Lecture-Tutorials* does require an investment in time, you do not have to figure out all of the nuances of the implementation strategy by yourself. By following the guidelines laid out in this chapter, you can avoid many of the mistakes we have made in the past or have seen others make. Taking the time to consider how your implementation is best suited to your institutional and classroom-specific context is critical. We strongly suggest that new and experienced faculty consider attending one of the many professional development workshops offered to help evolve faculty classroom practices, such as the Center for Astronomy Education's workshops (http://astronomy101.jpl.nasa.gov/workshops) and the Workshop for New Physics and Astronomy Faculty (http://www.aapt.org/Conferences/newfaculty/nfw.cfm). We find that faculty who take a scholarly and thoughtful approach to their teaching typically report experiencing greater enjoyment in their classroom, and pride in their students' dramatic gains in knowledge, and abilities.

**Acknowledgements**

The research and curriculum development efforts highlighted in this chapter have been supported by the National Science Foundation (Grant Nos. 9952232, 9907755, and 0715517) and by the generous contributions of NASA's Exoplanet Exploration Program and Associated Universities, Inc.